\documentclass[journal]{IEEEtran}
\usepackage{amsmath,amsfonts}
\usepackage{algorithmic}
\usepackage{algorithm}
\usepackage{array}
\usepackage[caption=false,font=normalsize,labelfont=sf,textfont=sf]{subfig}
\usepackage{textcomp}
\usepackage{stfloats}
\usepackage{url}
\usepackage{verbatim}
\usepackage{graphicx}
\usepackage{cite}
\usepackage{amsmath,amssymb,amsfonts}
\usepackage{graphicx}
\usepackage{textcomp}
\usepackage{xcolor}
\usepackage{siunitx}
\usepackage{paralist}
\usepackage{booktabs}
\usepackage{makecell} 
\def\BibTeX{{\rm B\kern-.05em{\sc i\kern-.025em b}\kern-.08em
    T\kern-.1667em\lower.7ex\hbox{E}\kern-.125emX}}
\begin{document}

\title{NVM-in-Cache: Repurposing Commodity 6T SRAM Cache into NVM Analog Processing-in-Memory Engine using a Novel Compute-on-Powerline Scheme }

\author{Subhradip Chakraborty, Ankur Singh, Xuming Chen,~\IEEEmembership{Graduate Student Member,~IEEE,} \\ Gourav Datta, and Akhilesh R. Jaiswal,~\IEEEmembership{Member,~IEEE}
        % <-this % stops a space
\thanks{Subhradip Chakraborty, Ankur Singh and Akhilesh Jaiswal are with the Department of Electrical and Computer Engineering, University of Wisconsin Madison, Madison, USA (e-mail: chakrabort42@wisc.edu, ankur.singh@wisc.edu).

Xuming Chen and Gourav Datta are with the Department of Electrical, Computer, and Systems Engineering, Case School of Engineering, Case Western Reserve University, USA.

%Subhradip Chakraborty and Ankur Singh contributed equally to this work.

}% <-this % stops a space
%\thanks{Manuscript received April 19, 2021; revised August 16, 2021.}
}

\markboth{NVM-in-Cache}%
{Shell \MakeLowercase{\textit{et al.}}: A Sample Article Using IEEEtran.cls for IEEE Journals}


\maketitle

\begin{abstract}
The rapid growth of deep neural network (DNN) workloads has significantly increased the demand for large-capacity on-chip SRAM in machine learning (ML) applications, with SRAM arrays now occupying a substantial fraction of the total die area. To address the dual challenges of storage density and computation efficiency, this paper proposes an NVM-in-Cache architecture that integrates resistive RAM (RRAM) devices into a conventional 6T-SRAM cell, forming a compact 6T-2R bit-cell. This hybrid cell enables Processing-in-Memory (PIM) mode, which performs massively parallel multiply-and-accumulate (MAC) operations directly on cache power lines while preserving stored cache data. By exploiting the intrinsic properties of the 6T-2R structure, the architecture achieves additional storage capability, high computational throughput without any bit-cell area overhead. Circuit- and array-level simulations in GlobalFoundries 22nm FDSOI technology demonstrate that the proposed design achieves a throughput of 0.4 TOPS and 452.34 TOPS/W. For 128 row-parallel operations, the CIFAR-10 classification is demonstrated by mapping a Resnet-18 neural network, achieving an accuracy of 91.76\%. These results highlight the potential of the NVM-in-Cache approach to serve as a scalable, energy-efficient computing method by re-purposing existing 6T SRAM cache architecture for next-generation AI accelerators and general purpose processors.
\end{abstract}

\begin{IEEEkeywords}
Analog in-memory computing, SRAM, 6T-Cell, NVM-in-cache, processing in-memory (PIM).
\end{IEEEkeywords}

\section{Introduction}
\IEEEPARstart{T}{he} growing demand for data-intensive computations, coupled with transistor scaling in accordance with the Moore’s law and the increasing complexity of consumer workloads, has driven steady growth in on-chip cache capacity over the years~\cite{cachesizetrend}. With the emergence of AI and deep neural networks, this trend has become even more critical to achieving state-of-the-art performance and precision. As the size of these networks increases, recent processors have required larger on-chip caches to meet performance needs~\cite{burd2022zen3}. Although several studies have explored cache architectures using 7T~\cite{7tsram}, 8T~\cite{8tsram}, 9T~\cite{9tsram}, and 10T~\cite{10tsram} SRAM cells, the conventional 6T SRAM still remains almost exclusively used in commercial processors due to its high density and fast access speed~\cite{tsmcsram}. However, with the rapid growth in computational workloads and the slowing benefits of technology scaling, there is a need to explore alternative memory technologies. Non-volatile memories (NVMs) offer smaller cell footprints, enabling higher density, and exhibit significantly lower leakage compared to conventional SRAM, which suffers from substantial leakage power due to short channel effects at advanced technology nodes~\cite{sramleak}. Some NVM options are Resistive RAM (RRAM)~\cite{rram}, Magnetic RAM (MRAM)~\cite{mram}, and Phase Change Memory (PCM)~\cite{pcm}. Despite their advantages, NVMs generally suffer from higher write latency and limited endurance, increased write power consumption, and lower signal-to-noise ratio (SNR) compared to SRAM~\cite{nvmprob}. Thereby, despite their density and power benefits, they are not suitable for on-chip high-speed cache applications. This motivates the integration of 6T SRAM with NVM devices, effectively combining the advantages of both NVM and 6T SRAM inside on-chip cache hierarchies, without incurring substantial (almost no) bit-cell area overhead.

In this work, we present an NVM-in-Cache architecture which integrates NVM within the 6T SRAM cells, thereby repurposing 6T SRAM to perform massively parallel analog processing-in-memory (PIM) operations using a novel compute-on-powerline scheme. Continuous technology scaling has enabled an increase in the number of processor cores on a chip, but memory bandwidth and energy efficiency have not scaled proportionally, leading to the well-known memory wall~\cite{memorywall}. In deep learning applications, the growing data transfer between memory and the processor has become a major bottleneck for energy-efficient computation. This has motivated the concept of PIM, also referred to as Computing-in-Memory (CIM), where computation is performed within the memory array to minimize data movement and improve efficiency in AI workloads. Prior NVM-based PIM designs have demonstrated this concept but often suffer from limited SNR ratio~\cite{snrprob}, low ON/OFF conductance ratio~\cite{lowonoff}, and fabrication challenges for large-scale integration~\cite{fabricationchallenge}. Among different NVM technologies, RRAM is notable for its relatively high conductance ratio~\cite{rram_ratio}, and has also been demonstrated as a scalable, high-density memory in commercial fabrication processes~\cite{rramden}. In this work we use RRAM for design and analysis, but the presented scheme is applicable to other NVMs as well. Note, while NVMs address some of the limitations of 6T SRAM, it is not an ideal direct replacement for cache memory due to its high write latency and energy consumption. This motivates the integration of NVM (RRAM) with conventional cache architectures, enabling efficient in-memory computation without compromising cache functionality.

\begin{figure*}[!t]
\centering
\includegraphics[width=1\linewidth]{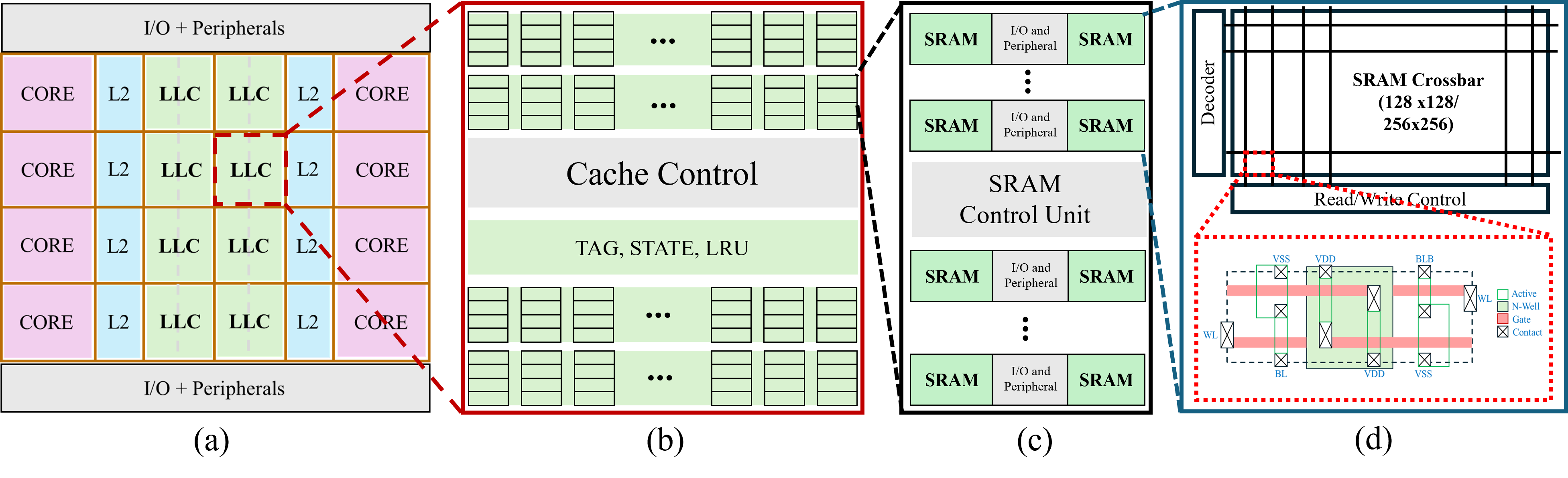}
\vspace{-0.3 in}
\caption{(a) Multi-core processor architecture consisting of multiple cores, private L2 caches, and distributed LLC slices; (b) Cache architecture of a single LLC slice, organized into multiple banks for parallel access; (c) Each bank is further divided into multiple smaller SRAM sub-arrays to improve scalability and reduce access latency; (d) A single sub-array block implemented using 6T SRAM cells, typically structured as a 128 $\times$ 128 or 256 $\times$ 256 array, along with its corresponding bit-cell layout.}
\vspace{-0.15 in}
\label{intro_cache}
\end{figure*}

Previous work on SRAM-based PIM~\cite{srampim1,srampim2,srampim3} has shown promising results but is generally not well-suited for cache architectures, as many designs employ more than six transistors per bit-cell, significantly reducing cache density. Some studies~\cite{6tsrampim1,6tsrampim2} have demonstrated PIM using standard 6T SRAM cells; however, these approaches require the cache to flush its existing data in order to store the neural network weights for in-memory computation. After computation, the cache data must be reloaded, which introduces additional latency and increases energy consumption due to extra data movement. 

%Additionally, during the sleep state of the SRAM, data can be backed up to an NVM to reduce reload operations, a capability that has been explored in prior work under the term NVSRAM~\cite{nvsram_jaydeep}. Our proposed solution retains this NVSRAM capability while enabling in-memory computation directly within the cache while preserving the original cache data, thereby eliminating the need for costly data flush and reload cycles.

In summary, the main contributions of this work are as follows:

\begin{itemize}
\item Novel 6T-2R bit-cell design that integrates resistive RAM (RRAM) devices with conventional 6T SRAM, enabling NVM in-memory analog computation (PIM mode).
\item Dual functionality of the proposed bit-cell, which doubles (NVM+SRAM storage in each bit-cell) the effective memory capacity of conventional cache architectures while maintaining minimal (almost no) bit-cell area overhead.
\item Processing-in-memory capability using a novel compute-on-powerline scheme, allowing NVM multiply-and-accumulate (MAC) operations to be performed directly within existing 6T SRAM cache arrays without modifying the cell footprint.
\item Massive array-level parallelism for analog PIM, enabling high-throughput computation while retaining the original cache data, thereby eliminating costly flush and reload operations.
\end{itemize}

Note, this work does not preclude implementing purely
SRAM-based IMC schemes (without NVM); rather, it can be
used in conjunction with such prior approaches, effectively
doubling the memory capacity and expanding the opportunities to realize PIM within conventional 6T cells. The remainder of the paper is organized as follows. Section II presents the preliminaries on RRAM and cache architectures. Section III details the various operational modes of the proposed 6T-2R bit-cell. Section IV provides an array-level analysis of the 6T-2R cache architecture. Section V presents simulation and validation results using commercial GlobalFoundries 22nm FDSOI technology. Finally, Section VI concludes the paper.

\section{Preliminaries \& Literature Overview}

\subsection{Resistive Random Access Memory (RRAM) Device}
RRAM is a type of resistive non-volatile memory that stores information based on the resistance state of the device with a high conductance ratio. It typically consists of silver (Ag) and platinum (Pt) electrodes separated by a conductive filament. The bipolar RRAM device operates with a positive SET voltage to switch to a low-resistance state (LRS) and a negative RESET voltage to return to a high-resistance state (HRS). The device switches from high- to low-resistance upon exceeding the SET voltage and returns to high-resistance when the RESET voltage is reached. A compact Verilog-A model that captures the non-linearity of the device and key parameters has been employed~\cite{rrammodel}. The LRS and HRS states represent the binary values 1 and 0, which are used to store neural network weights in our proposed scheme. These weights can be represented in an analog form either by physically adjusting the dimensions of the RRAM device or by tuning the programming states to achieve different resistance levels~\cite{rramden}. The latter approach is more flexible than altering the physical structure. The capability of RRAM is discussed in detail in Section 4.
\vspace{-0.1 in}

\begin{figure}[!t]
\centering
\includegraphics[width=1\linewidth]{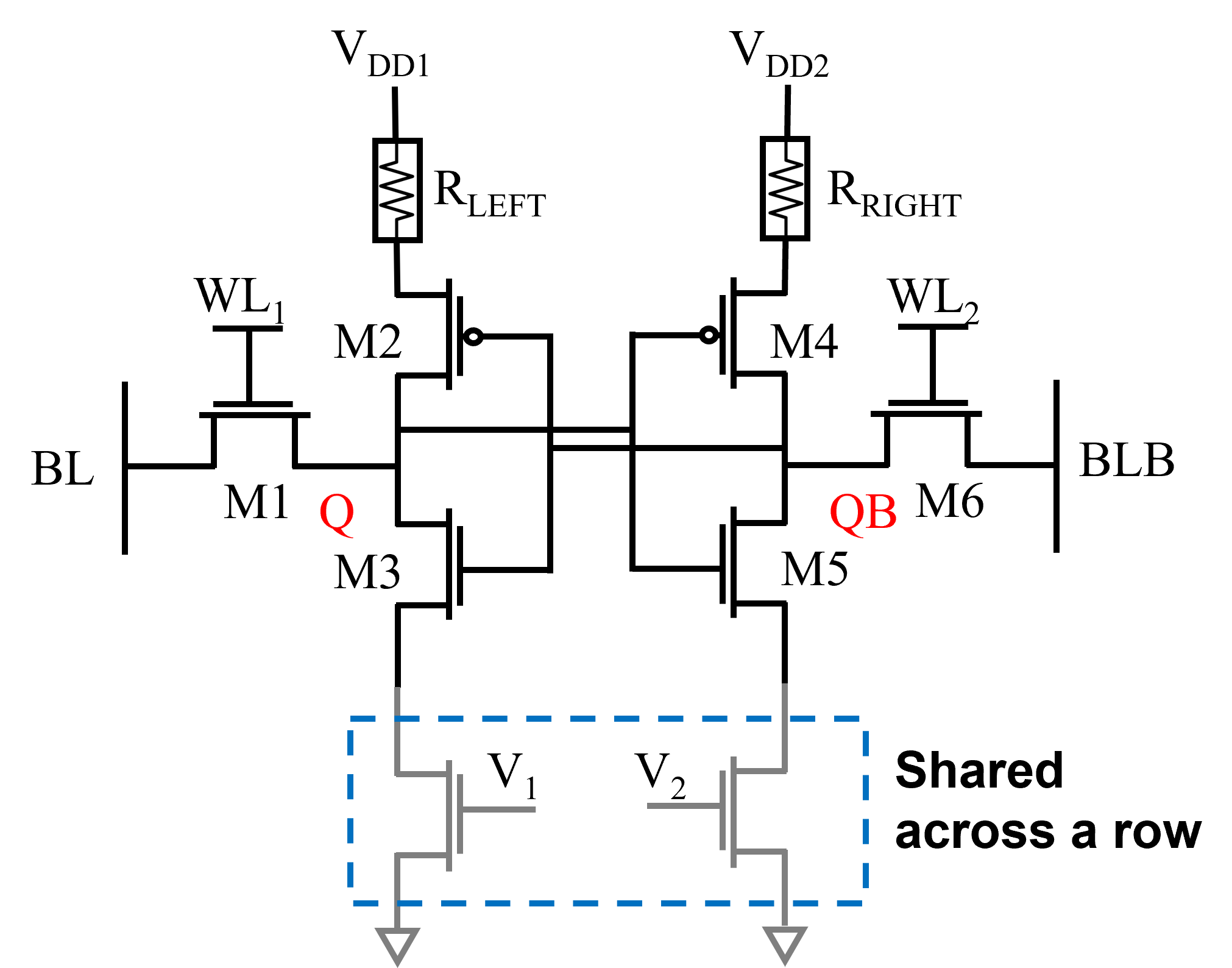}
\vspace{-0.2 in}
\caption{Propoed 6T-2R bit-cell along with the shared gated GND structure across a row.}
\vspace{-0.1 in}
\label{Proposed_bit-cell_intro}
\end{figure}

\begin{figure}[!t]
\centering
\includegraphics[width=1\linewidth]{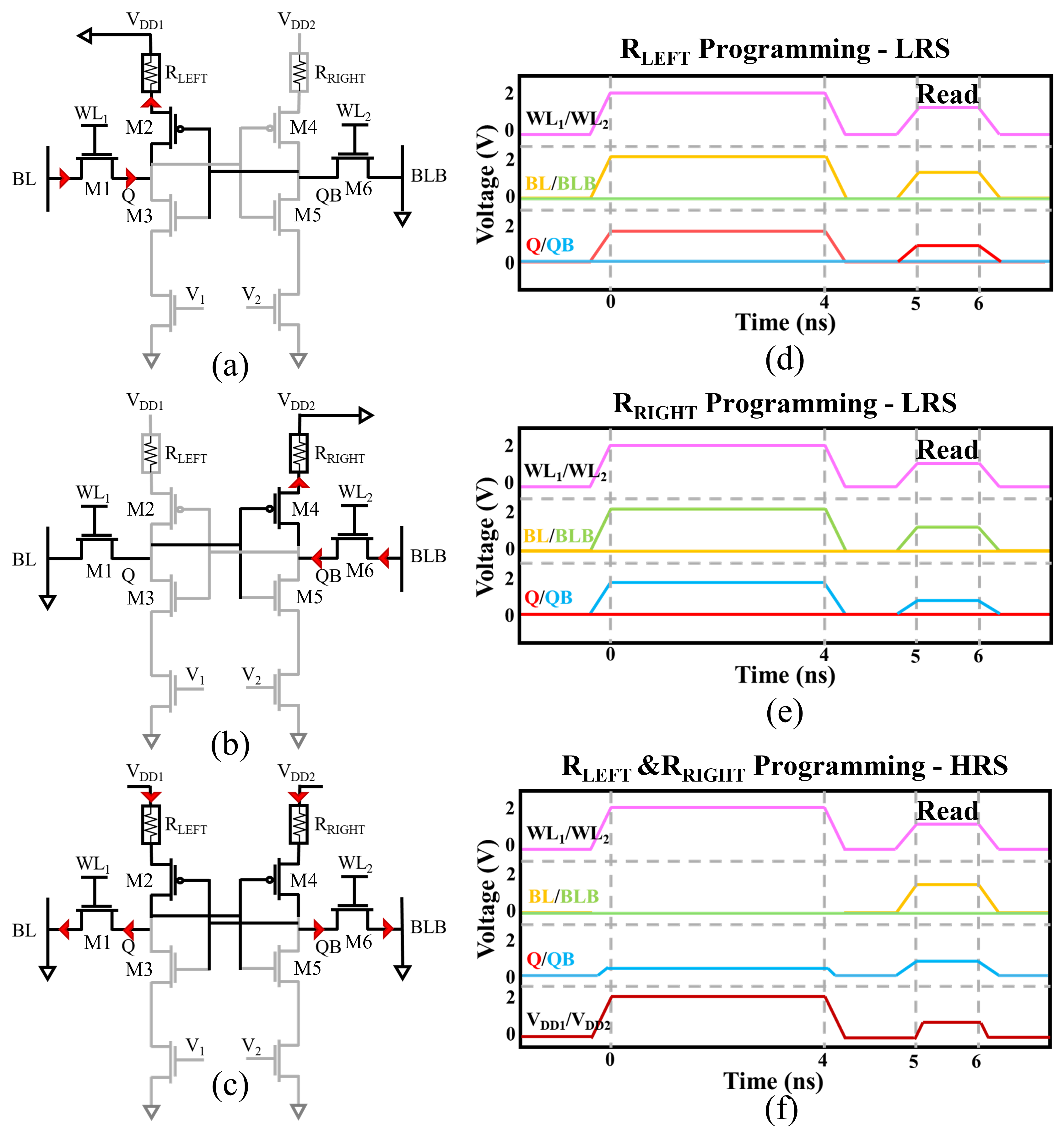}
\vspace{-0.3 in}
\caption{Programming operation of the proposed 6T-2R bitcell: (a) Current flows from BL to VDD1 while QB is driven to 0, turning ON PMOS M2 and programming $R_{LEFT}$ to LRS; (b) Current flows from BLB to VDD2 while Q is driven to 0, turning ON PMOS M4 and programming $R_{RIGHT}$ to LRS; (c) Current flows from VDD1 and VDD2 to BL and BLB, programming both $R_{LEFT}$ and $R_{RIGHT}$ to HRS. Timing diagrams of the control signals are shown for (d) programming $R_{LEFT}$ to LRS; (e) programming $R_{RIGHT}$ to LRS; and (f) programming both $R_{LEFT}$ and $R_{RIGHT}$ to HRS.
}
\vspace{-0.1 in}
\label{RRAM_prog}
\end{figure}

\subsection{Cache Architecture}
We provide a brief overview of the last-level cache (LLC) architecture by referencing the Intel multi-core Xeon and AMD Zen processor~\cite{intel1,intel2,neuralcache,amdzen5} as shown in Fig.~\ref{intro_cache}. The shared LLC is accessed by the multi-core processor via a ring/mesh interconnect. \cite{intel1} architecture consists of 8 LLC slices, each comprising a 20-way, 2.5 MB cache divided into 80 banks of 32 KB each. In addition to the data banks, each slice contains the corresponding 1-way tag, cache-valid (CV) bits, state, and least recently used (LRU) structures. Each 32 KB bank is further divided into 16 KB subarrays, and each subarray consists of two 8 KB (256 $\times$ 256) SRAM arrays, along with the necessary decoders, peripheral circuitry, and input/output interfaces. The focus of this work is to retain the overall architecture while replacing the conventional 6T SRAM cells with a 6T-2R structure. This modification not only preserves the standard cache functionality but also enables PIM capabilities, allowing computations to be performed directly within the memory arrays. By doing so, the proposed design leverages the benefits of NVM without compromising the cache’s original performance and organization.

\section{Design of the Proposed 6T-2R Bit-cell}

In this section, we describe the proposed 6T-2R bitcell in detail. A conventional 6T SRAM layout consists of ground lines (VSS), wordlines (WL), and specifically isolated power lines (VDD) which provides an opportunity to integrate non-volatile memory (NVM) between pull up devices and isolated VDD power lines. In our design the wordlines are separated (WL1 and WL2) which does not incur any additional area overhead~\cite{wlsep}. Further, two RRAM devices, $R_{LEFT}$ and $R_{RIGHT}$, are integrated along the VDD lines, while the VSS lines are extended across each row of the cache array to form a gated-GND structure~\cite{gatedvdd} as shown in Fig.~\ref{Proposed_bit-cell_intro}. This arrangement allows the 6T SRAM cell to be augmented with non-volatile memory devices in a seamless manner, thereby doubling the effective storage capacity without incurring any bit-cell area overhead. Beyond storage enhancement, the proposed structure also enables efficient support for PIM operations. The gated-GND control provides additional flexibility in managing the current flow through the cell, which is critical for ensuring reliable computation while retaining stored volatile data in the 6T cell. The bitcell itself is organized symmetrically, where the left and right halves are defined by powerlines (VDD1/VDD2), wordlines (WL1/WL2), bitlines (BL/BLB), and a dedicated shared control signal (V1/V2) that governs the gated-GND operation.

\vspace{-0.1 in}

\subsection{NVM Programming}
During the programming stage, only the 2T-2R portion of the 6T-2R bitcell, as highlighted in Fig.~\ref{RRAM_prog}, is utilized. Lets assume, the initial resistance state of the RRAM devices is set to HRS. To switch to LRS, a SET voltage must be applied, whereas a RESET voltage is required to transition back from HRS to LRS. The programming phase leverages the concept of wordline overdrive\cite{assisttech}. Programming from HRS to LRS requires two cycles, one for $R_{LEFT}$ and one for $R_{RIGHT}$, with each cycle having a pulse width of 4 ns. In the first cycle, $R_{LEFT}$ is programmed by overdriving WL1 and WL2 to 2 V, applying 2 V to BL and 0 V to BLB, and setting the gated-GND control signal (V1 and V2) to 0 V. Since the required SET voltage across the RRAM is 1.2 V, both VDD1 and VDD2 are grounded during this operation. In the second cycle, $R_{RIGHT}$ is programmed using the same method, except with BL = 0 V and BLB = 2 V, while the remaining signals are unchanged. During the LRS programming phase, the bitlines must be driven in complementary fashion to selectively activate the PMOS access transistors (M2 for $R_{LEFT}$ and M4 for $R_{RIGHT}$). This ensures a proper current path is established, enabling reliable programming of the RRAM devices.

Programming the RRAM devices to the HRS requires only a single cycle. During this operation, both wordlines (WL1 and WL2) are overdriven to 2 V, while BL and BLB are held at 0 V. The supply lines VDD1 and VDD2 are biased at 2 V, and the gated-GND control signal (V1 and V2) is set to 0 V. Unlike LRS programming, HRS programming does not require separate cycles for $R_{LEFT}$ and $R_{RIGHT}$, since both can be programmed in parallel. This is because, with BL and BLB grounded, both storage nodes (Q and QB) are also forced closed to 0 V, thereby turning on the PMOS transistors (M2 and M4) and enabling current flow from VDD1/VDD2 to BL/BLB. 

To validate successful programming, a read operation is performed immediately after the programming phase in a single cycle. For the read, BL and BLB are held at VDD (0.8 V), WL1 and WL2 are biased at VDD depending on whether to read $R_{LEFT}$ and $R_{RIGHT}$, and VDD1 and VDD2 are used to measure the current for 1ns. The current is then measured at the powerlines: if the RRAM device is in LRS, the measured current is high, whereas for HRS the current drops to low value, thereby confirming the programmed state. In our work, we use a programming voltage of 2 V, which is relatively high but determined by the specific RRAM model employed in this work. Use of RRAM devices optimized for low programming voltage will result in lesser overdrive. It is also important to note that programming is destructive to the SRAM data; however, in most neural network–based PIM applications, particularly for inference, the frequency of reads far outweighs programming operations, minimizing latency and power overhead. Furthermore, both $R_{LEFT}$ and $R_{RIGHT}$ are programmed to the same state (LRS or HRS) during PIM, which preserves the inherent symmetry of the SRAM cell.

\begin{figure}[!t]
\centering
\includegraphics[width=1\linewidth]{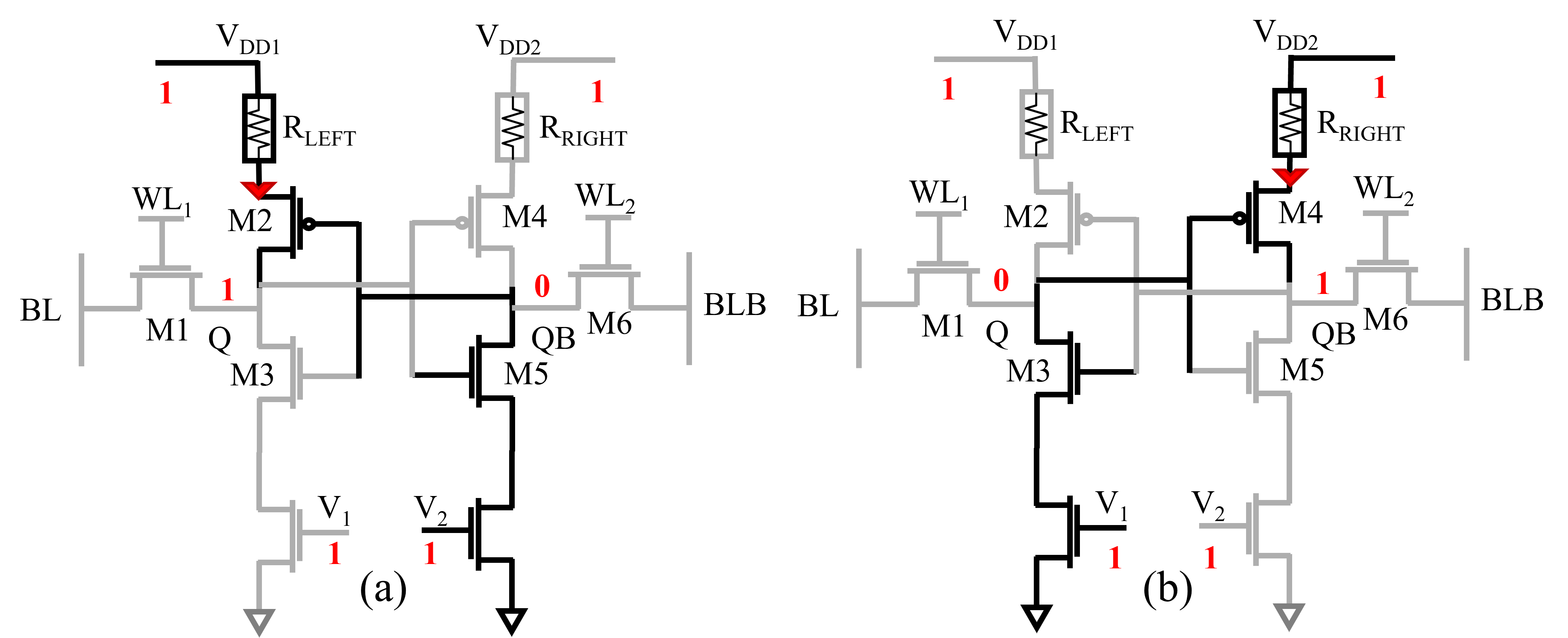}
\vspace{-0.1 in}
\caption{Hold operation in SRAM mode: (a) Q = 1 and QB = 0; (b) Q = 0 and QB = 1. The stored values are retained in a conventional SRAM manner, even with the placement of the RRAM devices on the power lines.}
\vspace{-0.1 in}
\label{SRAM_Hold_intro}
\end{figure}

\subsection{SRAM Hold/Read/Write Mode}
The proposed 6T-2R structure preserves the conventional read, write, and hold functionality of a standard 6T SRAM cache bit-cell. During the hold and read operation, the intrinsic advantage of placing the RRAM devices along the power lines becomes evident. Importantly, the resistance state of the RRAMs has no impact on data retention as shown in Fig.~\ref{SRAM_Hold_intro}. In the hold state, the supply voltages VDD1 and VDD2 are maintained at 0.8 V, which corresponds to the nominal operating voltage of the cell, while both wordlines (WL1 and WL2) are held low (0 V), V1 and V2 is set to 0.8 V as in a conventional 6T SRAM. Consider the case where the storage nodes Q/QB hold the values 1/0. Under this condition, PMOS transistor M2 remains ON, while M4 remains OFF. On the left half of the cell, the potential at node Q is equal to 0.8 V. Since there is no voltage difference between VDD1 (0.8 V) and node Q (0.8 V), no current flows through the path, thereby ensuring that Q remains strongly held at logic ‘1’ irrespective of the RRAM resistance. By the cross-coupled inverter structure, QB is correspondingly maintained at logic ‘0’. A similar argument applies when Q/QB = 0/1, in which case the complementary transistors switch roles to preserve the state. The hold operation is also tested on a 128 row 6T-2R array. This demonstrates that the addition of RRAM devices along the power lines does not interfere with the intrinsic bistability of the SRAM latch, and the stored data is reliably retained regardless of the resistance state of the RRAMs. The read and write operations of the proposed 6T-2R bitcell, when operating in SRAM mode, are identical to those of a conventional 6T SRAM. All control signals, including the wordlines (WL1 and WL2), bitlines (BL/BLB), and supply voltages, function in the same manner as in a standard SRAM cell~\cite{assisttech}.

%\vspace{-0.1 in}

\begin{figure}[!t]
\centering
\includegraphics[width=1\linewidth]{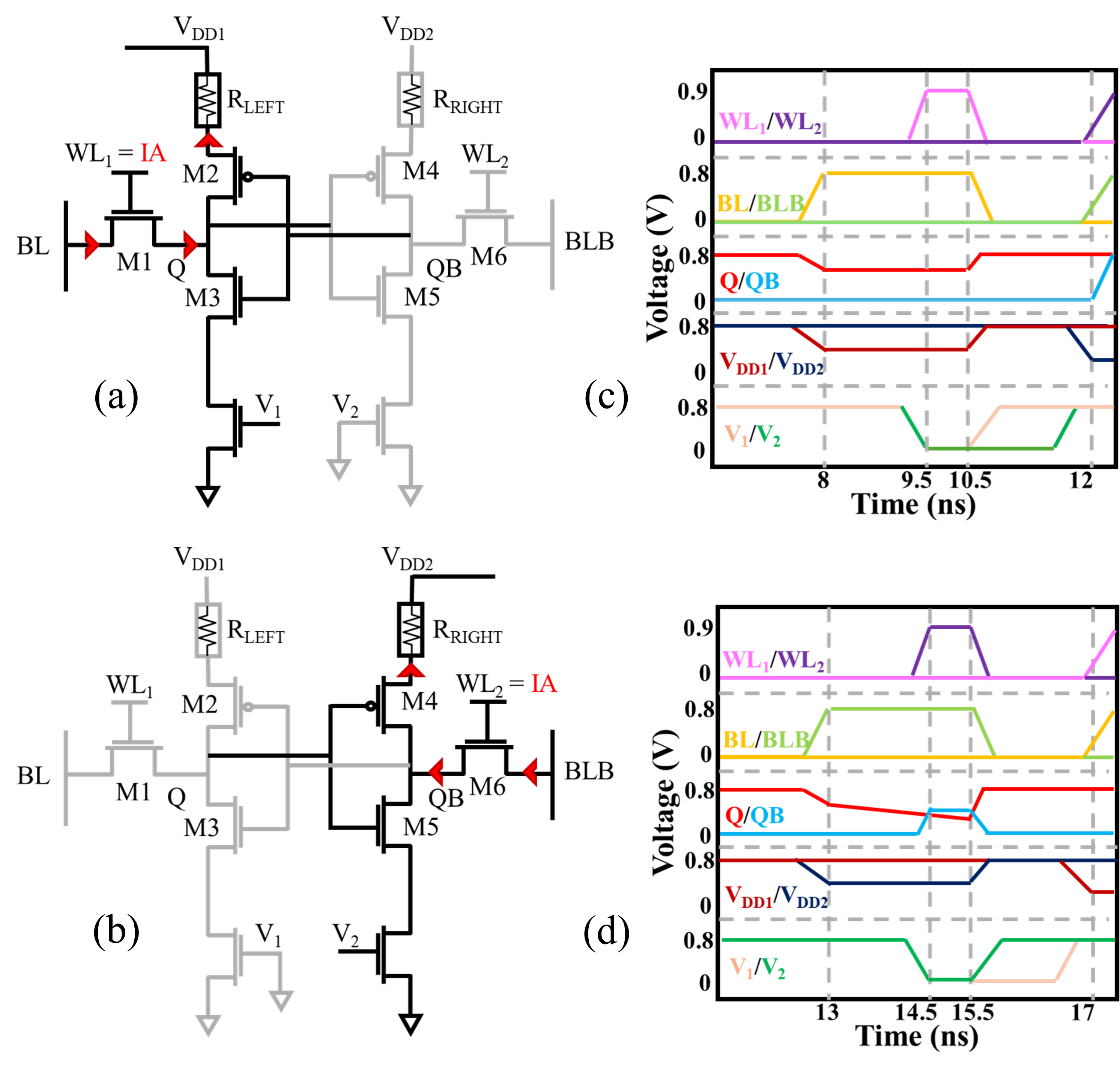}
\vspace{-0.2 in}
\caption{PIM dot-product operation in the proposed 6T-2R bitcell: (a) The operation is performed on the left half, where the weight is stored in $R_{LEFT}$ and the IA is applied to WL1, while the right half (QB) retains the cache data; (b) The same operation is performed on the right half. (c)-(d) Transient behaviour of the circuit for cache data Q = 1 and QB = 0.}
\vspace{-0.1 in}
\label{PIM_Mode_intro}
\end{figure}

\subsection{PIM Mode: Compute on Powerlines}
During PIM mode, the key idea behind the proposed 6T-2R bit-cell is to leverage the symmetry of the structure by assigning distinct roles to each half of the cell. For example, the left half of the bit-cell is utilized for performing dot-product computations, while the right half is reserved to reliably hold the existing cache data in the first cycle in a dynamic manner. This separation ensures that computation can be carried out without disturbing the stored cache values. %Before initiating conventional cache operations, the RRAM devices must first be programmed to represent the weights of the neural network, as explained in previous subsections.

Let us assume the RRAM devices have been programmed to the desired state corresponding to the weights of a deep learning network. The PIM phase of operation corresponds to the in-memory dot-product computation, which is executed in two cycles, each lasting 3.5 ns. The core idea is to exploit the programmed resistance states of the RRAM devices as weights and the applied wordline values as input activations (IA), while the resulting output current is collected along the power lines (VDD1/VDD2). Since VDD1 and VDD2 are shared across a column of the cache array, the output currents from multiple cells naturally accumulate in the current domain, thereby enabling efficient MAC operations. 

In the first cycle, WL1 and WL2 are held at 0 V for the initial 1.5 ns, while BL is driven to 0.8 V. During this period, the gated-GND control signals V1 and V2 are maintained at 0.8 V. Simultaneously, VDD1 is switched from 0.8 V to the reference voltage provided by the weighted configuration circuit (WCC), which performs the weighted current addition, while VDD2 remains at 0.8 V. For the subsequent 1 ns, WL1 is driven by the IA, V1 and V2 are pulled to 0, and the output current is measured at the VDD1 line. This is followed by 1 ns during which VDD1 and V1 are restored to the nominal value of 0.8 V. Finally, V2 is also returned to 0.8 V, restoring SRAM hold. The second cycle follows a similar sequence. WL1 and WL2 are set to 0 V for the first 1.5 ns, and BLB is driven to 0.8 V. V1 and V2 remain at 0.8 V during this period, while VDD2 is switched from 0.8 V to the bias voltage provided by the WCC, with VDD1 held constant at 0.8 V. In the next 1 ns, WL2 is applied with the IA, V1 and V2 are pulled to 0, and the output current is sampled at the VDD2 line. This is followed by 1 ns during which VDD2 and V2 are restored to 0.8 V. Finally, V1 is also returned to 0.8 V, completing the cycle and re-establishing SRAM hold. The output current through VDD1 and VDD2 are sampled only when IA is applied at WL1 and WL2, thereby performing analog dot product operation.

Consider the case where the stored values at the internal nodes are Q = 1 and QB = 0, as shown in Fig.~\ref{PIM_Mode_intro}. Under this condition, the PMOS transistor M2 is switched ON, while M4 remains OFF. During the first 1.5 ns (starting at 8 ns in Fig.~\ref{PIM_Mode_intro}(b)), VDD1 is pulled to a reference voltage provided by the WCC, which simultaneously discharges Q from 1 (0.8 V) to the reference voltage while keeping WL1 = 0 and VDD2 = 0.8 V. This interval of 1.5 ns is used as a precaution since VDD1 is shared across 128 cells along a column, and the parasitic capacitances require time to settle. In the subsequent 1 ns, if WL = 1 (IA = 1), the programmed resistance state of $R_{LEFT}$ determines the current magnitude: either a large current (LRS) or a small current (HRS) flows through VDD1. With V2 disabled (0 V), QB is held at 0 V, thereby ensuring data retention during computation. In the following 1 ns, both V1 and VDD1 are restored to 0.8 V. Since M2 remains ON (as QB = 0), Q recharges to 0.8 V, completing the cycle. During the second cycle, for the first 1.5 ns, Q remains at logic 1, which keeps M4 OFF and QB at 0. In the next 1 ns, WL2 = 1 and BLB = 1, charging QB to logic 1. In this case, NMOS transistor M3 turns ON, but since V1 = 0 V, Q is not discharged via M3. With M4 OFF, only minimal current is observed at VDD2. In the following 1 ns, when V2 is restored to 0.8 V, NMOS M5 switches ON and discharges QB, thereby preserving the original data. After completing the computation cycle, the circuit transitions back to the SRAM hold state.

For IA = 0, during the first 1.5 ns, Q discharges from 0.8 V to the reference voltage. Consequently, during the sampling window (when IA = 0 is applied to WL1), there is no voltage difference between Q and VDD1, resulting in negligible current flow through the power lines and ensuring a dot-product result of 0. Thus, when Q = 1 and QB = 0, the dot-product output is obtained on the left-hand side, whereas when Q = 0 and QB = 1, the output is obtained on the right-hand side. In this way, the two-cycle operation enables dot-product computation regardless of the data stored within the SRAM cell. This approach allows the bit-cell to alternate seamlessly between conventional SRAM functionality and in-memory MAC operations, while preserving data integrity.

\begin{figure*}[!t]
\centering
\includegraphics[width=1\linewidth]{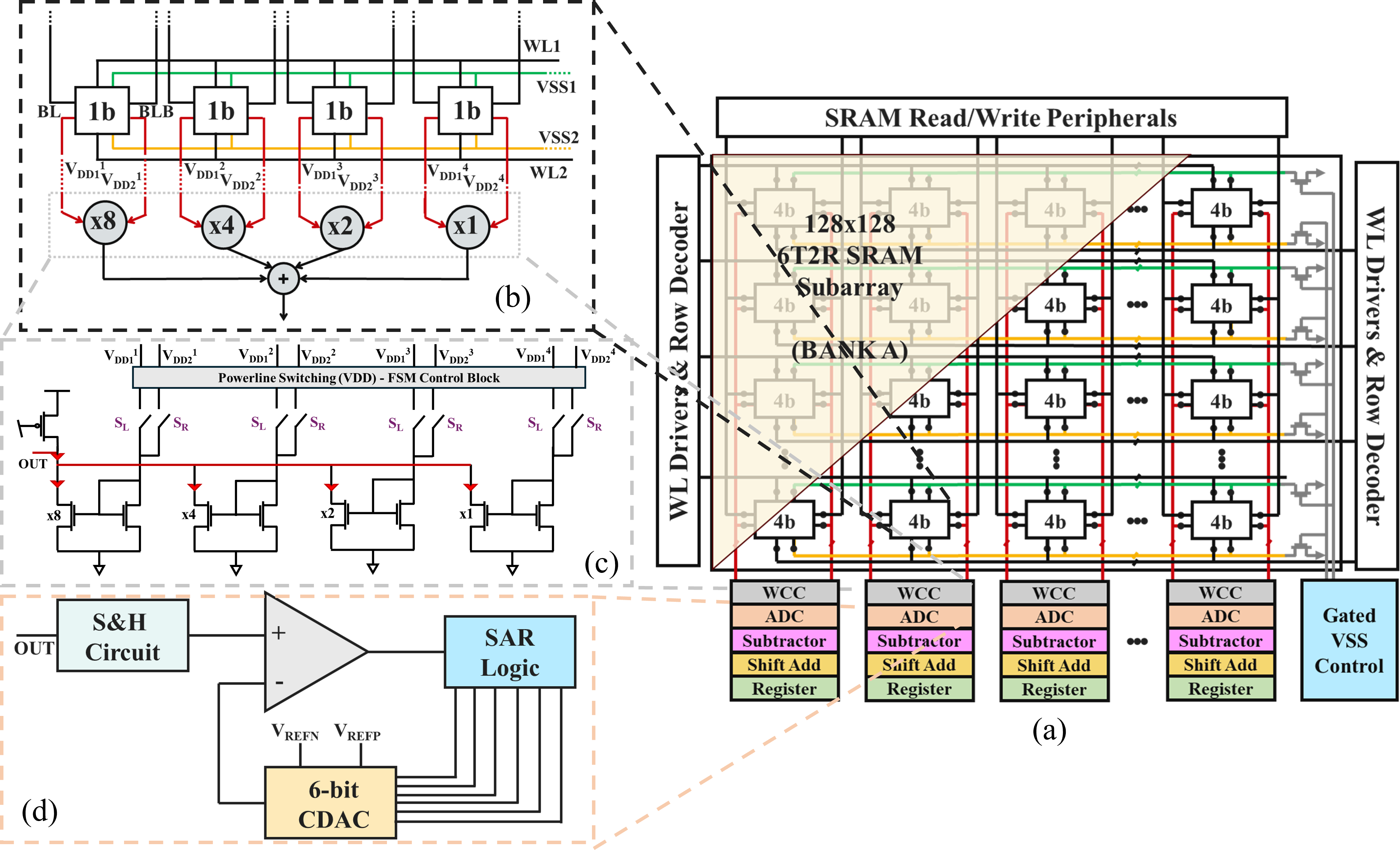}
\vspace{-0.1 in}
\caption{Array-level architecture: (a) Overview of the 128 $\times$ 512 6T-2R sub-array; (b) 4-bit word of the 6T-2R sub-array, where the current is collected from the VDD lines during PIM mode and sent to the WCC unit for weighted summation; (c) Block diagram of the analog WCC consisting of current mirror circuits to sample the weighted current; (d) The weighted current is passed to a 6-bit SAR ADC for digital conversion.}
\vspace{-0.1 in}
\label{array}
\end{figure*}

\section{Array Level Analysis}

\subsection{Massive Parallel MAC Operation}
The proposed 6T-2R bit-cell is organized in a banking structure, similar to those employed in the last-level caches of conventional processor architectures. Each unit subarray is designed as an 8 KB block as shown in Fig.~\ref{array}, consisting of a $128 \times 128$ SRAM array with 4-bit words, which can be equivalently viewed as a $128 \times 512$ 1-bit cache structure. In this organization, VSS lines are shared along the rows together with the wordlines, while VDD lines are shared along the columns along with the bitlines (BL/BLB). During PIM operation, IA are applied across the rows via WL1 and WL2 in two cycles, while the weights are stored in the RRAM devices embedded within each 6T-2R bit-cell. Consequently, each bit-cell performs a dot-product operation, and partial sums are accumulated along the shared VDD lines, enabling simultaneous accumulation from 128 bit-cells. This MAC operation is inherently parallel across columns as well, thereby allowing the 6T-2R architecture to exploit both row- and column-level parallelism. 

%Unlike conventional NVM designs that are constrained by the number of row activations, this structure supports fully parallel computation across the entire array, thereby enabling massive-scale parallel MAC operations.

%\vspace{-0.1 in}

\subsection{Multi-bit PIM Computing}
To enable multi-bit computation within the proposed 6T-2R cache-based architecture, we implement a $4\text{b} \times 4\text{b}$ MAC operation between the stored weights and IA for a 128-row activation simultaneously. Each 4-bit word corresponds to the weight storage, while the input activation is shared across the rows. The computation is carried out using a bit-serial scheme, which requires four cycles to complete the multi-bit operation. Although a bit-parallel approach using a digital-to-analog converter (DAC) could in principle execute the multi-bit IA in a single cycle, such an implementation would introduce significant peripheral overhead in conventional cache architectures. Specifically, it would also increase the design complexity required for accurate quantization of the accumulated current from the MAC array. For this reason, we adopt the bit-serial approach as a more area- and power-efficient strategy~\cite{bitserial}.

Within each 4-bit word, the outputs are distributed across eight VDD lines (four from the left, denoted VDD1, and four from the right, denoted VDD2). These signals are routed to the weighted configuration circuit (WCC), which applies an 8:4:2:1 weighting to represent the most-significant to least-significant bits (MSB–LSB) of the 4-bit weights. The WCC is controlled by a shared finite-state machine (FSM) that dynamically switches the VDD lines between a nominal 0.8 V mode and PIM mode during the sampling phase, before restoring them to 0.8 V once sampling is complete, as described in the previous section. The currents obtained at the VDD lines are processed through an NMOS current mirror, which scales the contributions according to the 8:4:2:1 ratio and combines them in the current domain. The accumulated current enables sampling onto a capacitor inside the sample and hold circuit as shown in Fig.~\ref{array}(d).

The sampled output voltage across the capacitor is then digitized using a 6-bit successive-approximation register (SAR) ADC. The ADC operates at 50 MHz and employs a strong-arm latch comparator, together with a 6-bit capacitive DAC and SAR logic implementing the binary search algorithm. This conversion process is performed twice, once for the left-side outputs (VDD1) and once for the right-side outputs (VDD2). In the post-processing stage, the final ADC output is inverted with respect to the MAC value, since during sampling the current in the WCC flows from VDD. As a result, the output voltage shown in Fig.~\ref{array}(c) corresponds to VDD – MAC. Following digitization, the outputs are processed through a sequence of digital circuits, including shift-and-add units, subtractors, and output registers, to generate the final accumulated result. These digital operations can be implemented outside the cache array.

%\vspace{-0.1 in}

\begin{figure}[!t]
\centering
\includegraphics[width=1\linewidth]{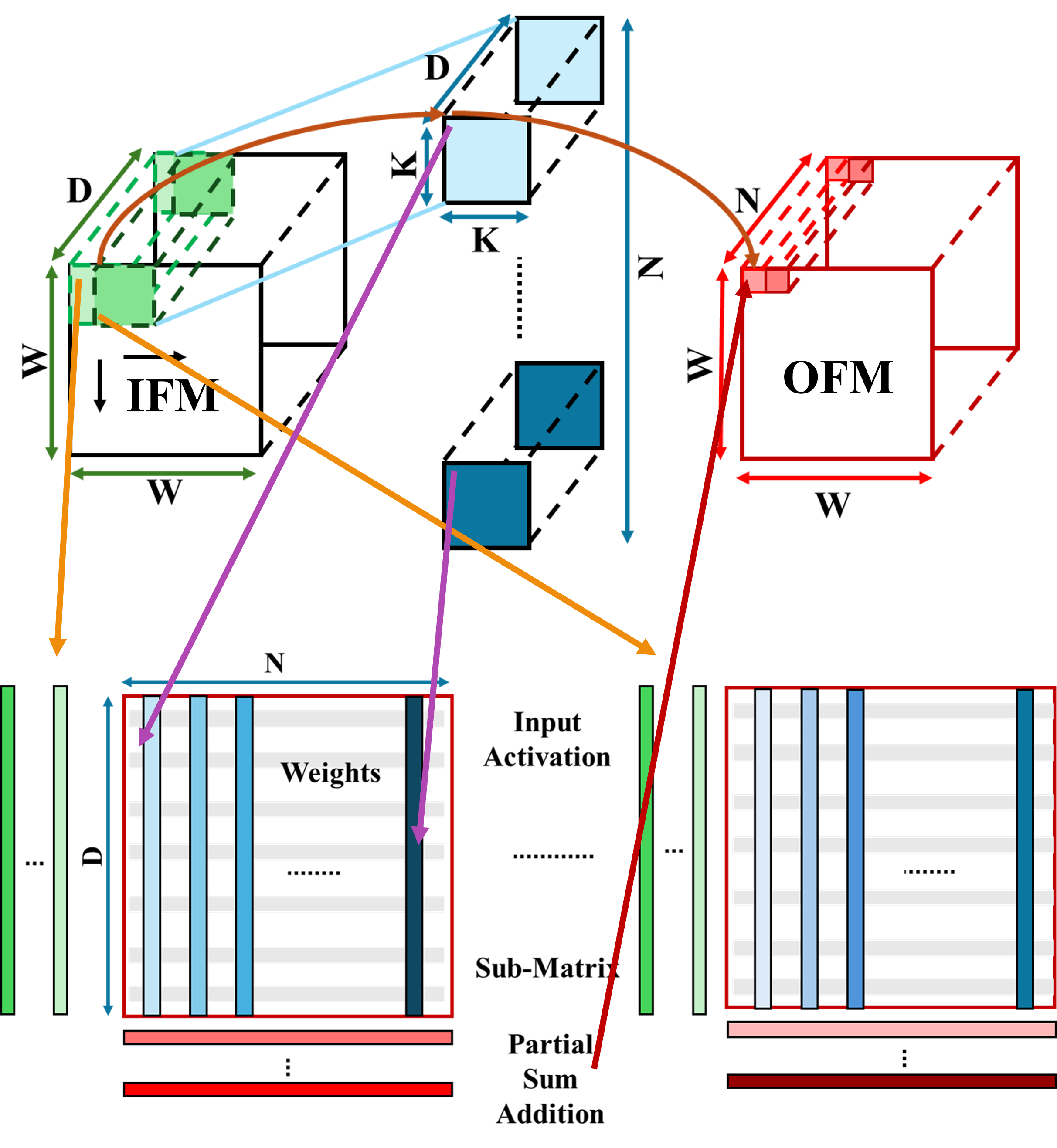}
\vspace{-0.2 in}
\caption{\textcolor{blue}{Input reuse mapping technique that distributes and maps kernel weights across the sub-arrays for efficient computation considering a stride of 1.}}
\vspace{-0.1 in}
\label{Mapping}
\end{figure}

\begin{figure*}[!t]
\centering
\includegraphics[width=1\linewidth]{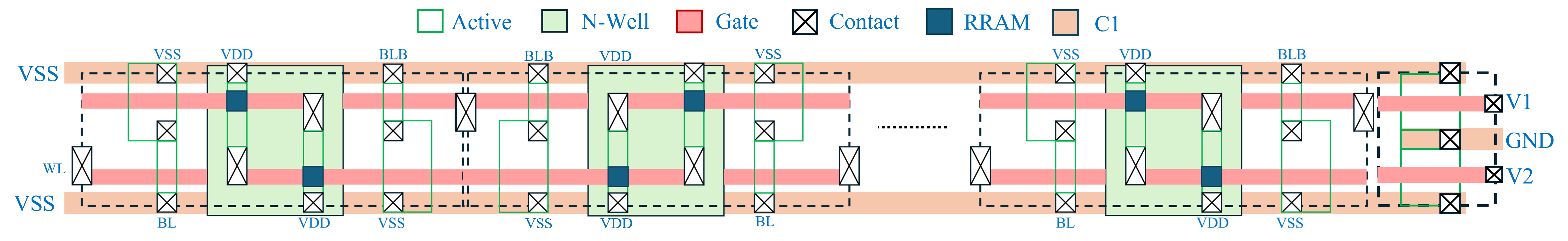}
\vspace{-0.2 in}
\caption{Layout of a single row of the 6T-2R bitcell with the gated GND structure, designed using the GlobalFoundries 22nm FDSOI PDK.}
\vspace{-0.1 in}
\label{Layout}
\end{figure*}

\subsection{Mapping of Weights and Input Activations}
We take the case of convolutional neural networks (CNNs), in CNNs, the size of the output layer is determined by convolving the input feature map (IFM) with the 3D weight kernel. The IFM has dimensions $W \times W \times D$, where $W$ denotes the input image width and $D$ represents the depth of the feature map. The kernel dimensions are $K \times K \times D \times N$, where $N$ is the number of output features. With same padding and a stride of 1, the resulting output feature map (OFM) has dimensions $W \times W \times N$ depending on the stride value. Convolution is performed by sliding the kernels across the IFM, carrying out element-wise multiplications followed by accumulation to generate the OFM.

We adopt the IFM reuse mapping strategy described in~\cite{mappingiscas}. In the proposed 6T–2R architecture, the IA are applied along the rows (wordlines, WL1 and WL2), while the outputs are also collected across the columns (VDD1 and VDD2). The weights are stored in the 6T–2R array, as illustrated in Fig.~\ref{Mapping}, while the IFMs are mapped column-wise according to the kernel dimensions. Specifically, $K \times K$ submatrices are generated, and the corresponding input data for each kernel position is directed to its associated submatrix. Each submatrix is mapped onto a 128 $\times$ 128 6T–2R bank (sub-array). During the first cycle, IFMs are multiplied with the 3D kernels mapped across the banks, producing partial sums that are accumulated outside the array to generate the initial OFMs. In the following cycle, the IFMs shift by one stride. Importantly, IFMs from the first cycle are reused in the second cycle as neighboring banks forward the required IFMs. Repeating this process over multiple cycles produces the final OFM.

When the application demands higher input or weight bit precision than supported by the hardware, multiple column outputs can be shifted and added in the digital domain using the shift-and-add block. To handle both positive and negative weights, separate memory banks are designated for each. The outputs from these banks are then combined in the digital domain using the subtractor block. 

%A key design choice in this work is the use of 128 $\times$ 128 banks instead of larger 256 $\times$ 256 banks. For the initial layers of CNNs, the feature map depth is relatively small, and larger arrays result in poor utilization of compute resources. Conversely, reducing the bank size below 128 introduces additional area overhead due to the ADC units, which in turn reduces compute density and also requires multiple cycles to process deeper layers. The 128 $\times$ 128 configuration thus represents a balanced design trade-off between utilization efficiency, area, and performance.

\section{Results \& Discussion}

\subsection{Bit-cell and Word Layout}
The layout study is conducted using the GlobalFoundries 22 nm FDSOI PDK. The standard 6T SRAM cell occupies an area of around 0.1 um\textsuperscript{2}~\cite{gfsram}, with wordlines routed separately, VSS routed horizontally, and VDD routed vertically. Within the bit-cell, the RRAM devices are symmetrically placed on top of the PMOS source terminals and the VDD lines. A 4-bit word is formed by connecting four such SRAM bit-cells, where the VSS lines are shared row-wise and subsequently connected to the gated GND transistors (V1 and V2), as illustrated in Fig.~\ref{Layout}.

Additionally, if VDD is routed horizontally and VSS is routed vertically, the IA could be applied on the bitlines (BL and BLB) instead of the wordlines (WL1 and WL2). The same process could then be used to perform the PIM operation. However, to remain consistent with the standard 6T SRAM layout, we chose to apply the IA on the wordlines.

\begin{figure}[!t]
\centering
\includegraphics[width=1\linewidth]{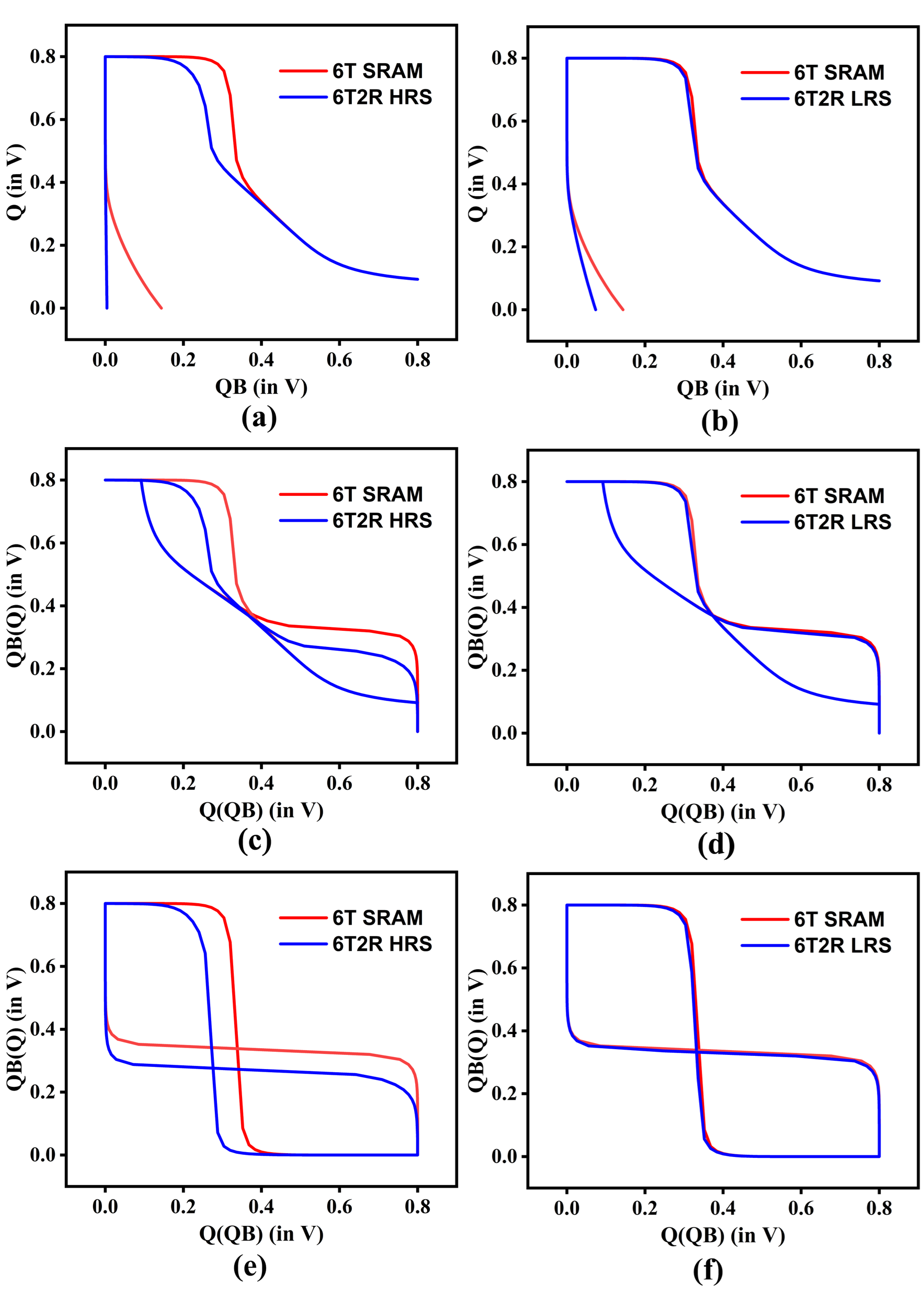}
\vspace{-0.3 in}
\caption{Comparison of conventional 6T SRAM and proposed 6T-2R bitcell: (a)-(b) Write Hold, (c)-(d) Read, and (e)-(f) Hold Static Noise Margin (SNM).}
\vspace{-0.1 in}
\label{rram_curve}
\end{figure}

%\begin{table}[!t]
%\centering
%\caption{Comparison of Restore Energy and Delay with Prior Works}
%\begin{tabular}{|c|c|c|}
%\hline
%\textbf{Ref.} & \textbf{Restore Energy} & \textbf{Restore Delay} \\ \hline
%9T\cite{tripathi2022novel}   & 15.20 fJ & 52.0 ps  \\ \hline
%10T\cite{tripathi2022novel}  & 16.54 fJ & 55.9 ps  \\ \hline
%9T\cite{wang2020magnetic}   & 15.34 fJ & 53.1 ps  \\ \hline
%10T\cite{wang2020magnetic}  & 16.77 fJ & 56.4 ps  \\ \hline
%8T\cite{tripathi20238t}   & 3.816 fJ & 228.8 ps \\ \hline
%\textbf{This Work} & \textbf{1.865 fJ} & \textbf{176 ps} \\ \hline
%\end{tabular}
%\label{tab:restore_comparison}
%\end{table}

\vspace{-0.1 in}
\subsection{6T-2R Array Level Performance}

The resistive switching behavior of the bipolar RRAM device is illustrated in Fig.~\ref{rram_curve} (a), showing the typical I--V hysteresis curve. Specifically, the device undergoes a \textit{SET} transition when the applied voltage exceeds the positive threshold ($V_{\text{set}} = +1.2~\text{V}$), switching from the high-resistance state (HRS, $\sim 300~\text{k}\Omega$) to the low-resistance state (LRS, $\sim 6.25~\text{k}\Omega$). Conversely, the \textit{RESET} transition occurs at a negative threshold voltage ($V_{\text{reset}} = -1.2~\text{V}$), returning the device to HRS. These transitions are highly reproducible and clearly demonstrate the non-volatile nature of the RRAM, where the resistance state is preserved even after removal of the bias voltage. Furthermore, in the write operation, the device requires only $4~\text{ns}$ for successful programming (SET or RESET). The high ratio between HRS and LRS ensures a clear distinction between logic ``0'' and ``1,'' making the device highly suitable for binary weight storage. In our simulations, a $0.8$--$1.05~\text{V}$ read voltage was used for a $1-2~\text{ns}$ read window, which is sufficient to measure the conductance without altering the memory state.

We next examine the robustness of the proposed design in terms of static noise margins (SNM). Fig.~\ref{rram_curve} presents a comparative analysis of read, write, and hold SNMs for the proposed 6T-2R bit-cell for different resistance state against a conventional 6T SRAM. Despite the addition of two RRAM elements and the gated GND control transistors, the proposed design exhibits only marginal changes in the noise margins compared to the baseline 6T cell. During the hold operation, the butterfly curve shows that the proposed bit-cell retains nearly identical static noise margins as the conventional design, thereby confirming that the integration of non-volatile elements does not compromise the bistability of the cross-coupled inverters. In the read mode, the proposed cell demonstrates a slight reduction in SNM compared to the 6T SRAM, due to the additional series resistance introduced by the RRAM devices. However, this degradation remains minimal and well within acceptable limits for reliable operation which could be enhanced by read-assist technique. %Furthermore, the gated GND scheme helps mitigate read-disturb issues by dynamically biasing the pull-down paths, ensuring that the stored state remains unaffected during wordline activation. 
Although the write SNM is somewhat reduced compared to the 6T SRAM, this trade-off is acceptable, and writability can be further enhanced through conventional write-assist techniques. The read latency of the baseline 6T SRAM is $660~\text{ps}$, while the proposed 6T-2R incurs a slightly higher latency of $686~\text{ps}$, reflecting minimal impact of the altered bit-cell. This small overhead is a reasonable trade-off for the additional non-volatile and in-memory computing functionality.

%In addition to static stability, the dynamic performance of the proposed 6T-2R NV-SRAM was thoroughly evaluated.  Beyond read operations, the NV-SRAM mode was characterized in terms of restore performance. The restore latency was measured to be $176~\text{ps}$, while the restore energy for a $512$-bit row was $0.955~\text{pJ}$. As shown in Fig.~\ref{restore}, both restore delay and energy remain stable across supply voltage variations (Fig.~\ref{restore}a) and process corners (Fig.~\ref{restore}b). Furthermore, a comparison with prior works is provided in Table~\ref{tab:restore_comparison}, where the proposed design achieves the lowest restore energy (1.865~fJ/bit) while maintaining competitive delay (176~ps), thereby demonstrating its superiority in energy-efficient non-volatile SRAM design.

\vspace{-0.1 in}

\begin{figure}[!t]
\centering
\includegraphics[width=1\linewidth]{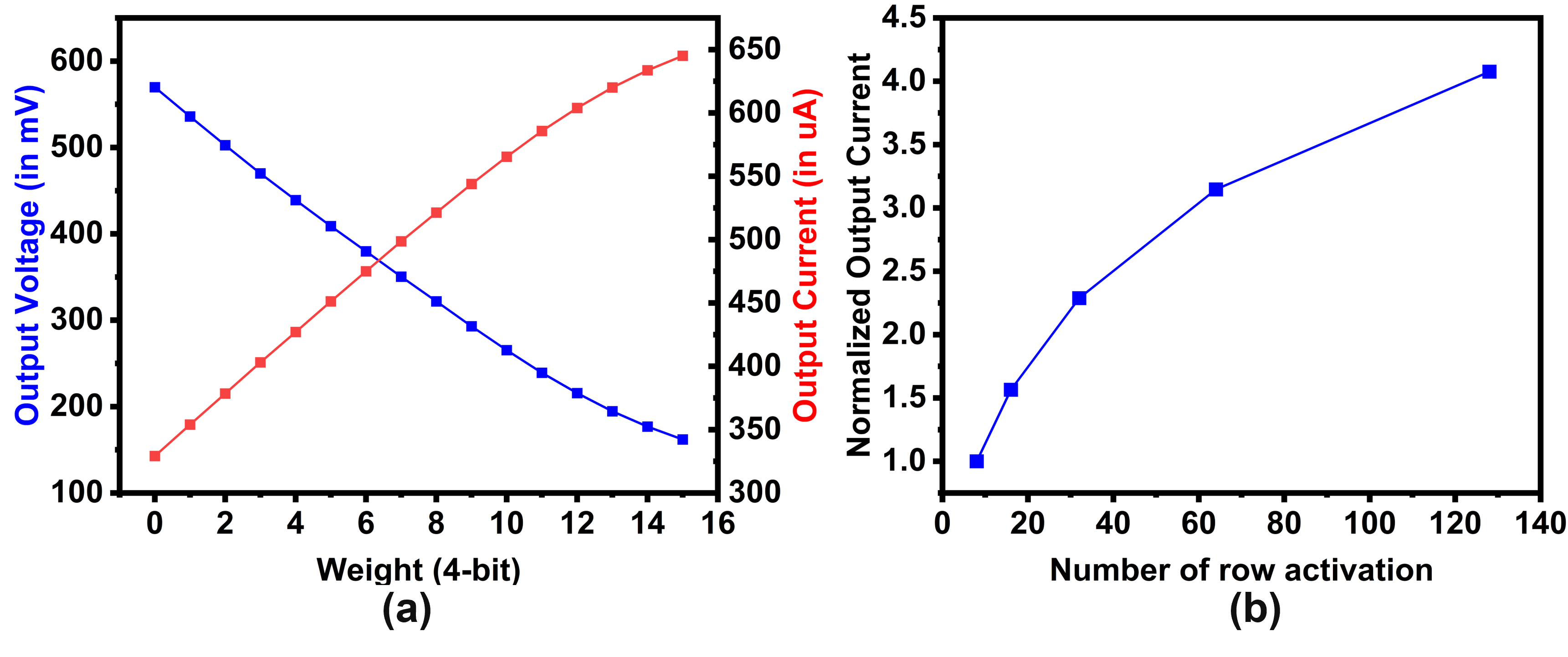}
\vspace{-0.3 in}
\caption{Linearity characteristics of the proposed 6T-2R bit-cell across process corners for 128 row activation: (a) weight vs. accumulated output voltage and current, and (b) normalized output current vs number of row activations.}
\vspace{-0.1 in}
\label{corner_linear}
\end{figure}

\begin{figure}[!t]
\centering
\includegraphics[width=1\linewidth]{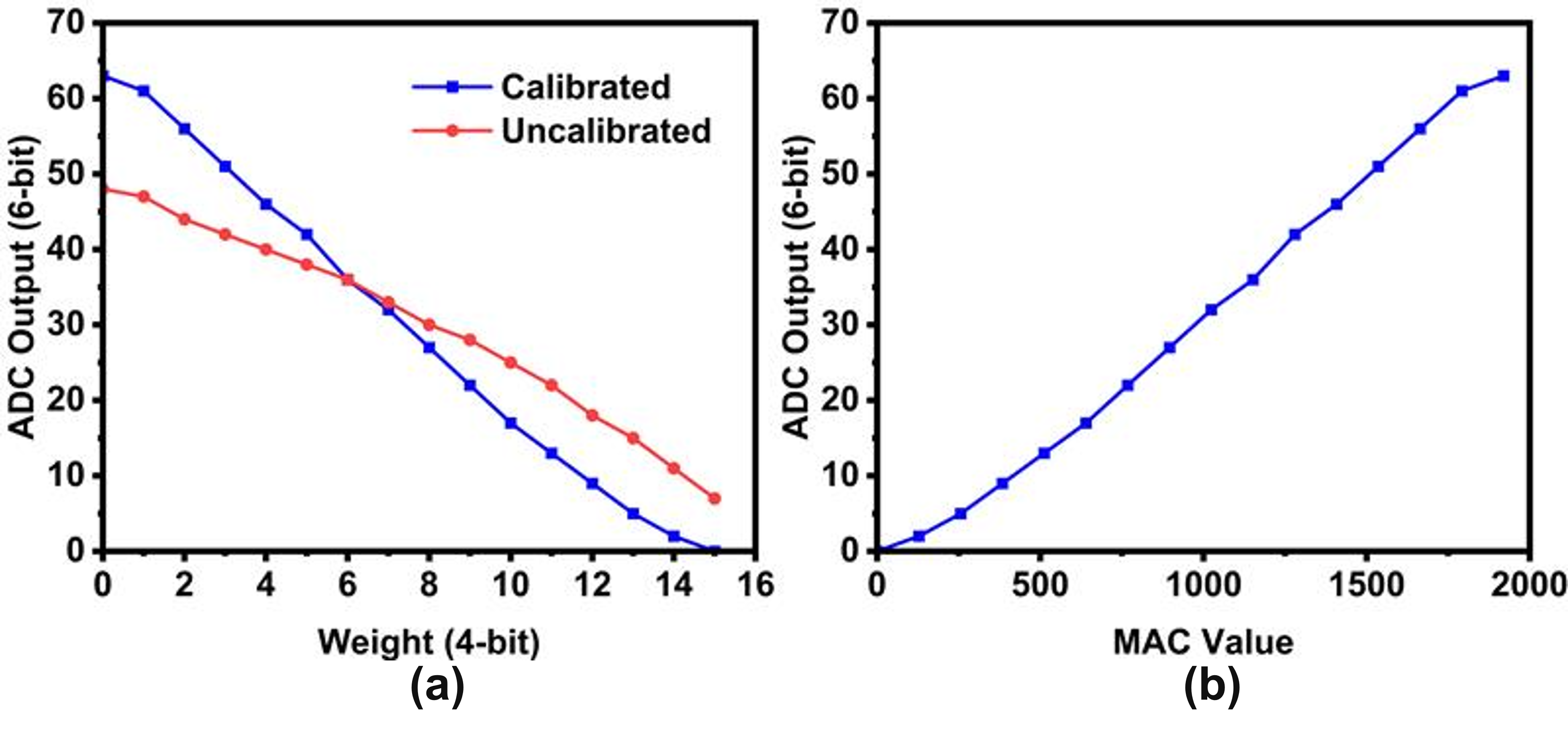}
\vspace{-0.3 in}
\caption{ADC characterization of the proposed 6T-2R PIM for 128 row activation: (a) weight-to-ADC output transfer curves with calibrated ($V_{\text{REFP}}=570$~mV, $V_{\text{REFN}}=155$~mV) and uncalibrated ($V_{\text{REF}}=800$~mV) SAR ADC, and (b) ADC output response as a function of accumulated MAC value.}
%\vspace{-0.1 in}
\label{adc_linearity}
\end{figure}

\subsection{Linearity Analysis}
The linearity of the proposed 6T-2R bit-cell was evaluated by sweeping the programmed weight ($w=0$--$15$) and recording both the accumulated column current and the corresponding sense voltage. Fig.~\ref{corner_linear}(a) shows the weight-to-voltage characteristics after sampling. Fig.~\ref{corner_linear}(b) showcase the normalized output current output vs number of row activation. The amount of nonlinearity is attributed to the stronger transistor drive at different corners, which reduces the effective voltage swing across the RRAM stack and limits incremental current contributions, but this nonlinearity can be accounted for while training the neural network model. Monotonicity is preserved across all corners, ensuring correct weight ordering.

\begin{figure}[!t]
\centering
\includegraphics[width=1\linewidth]{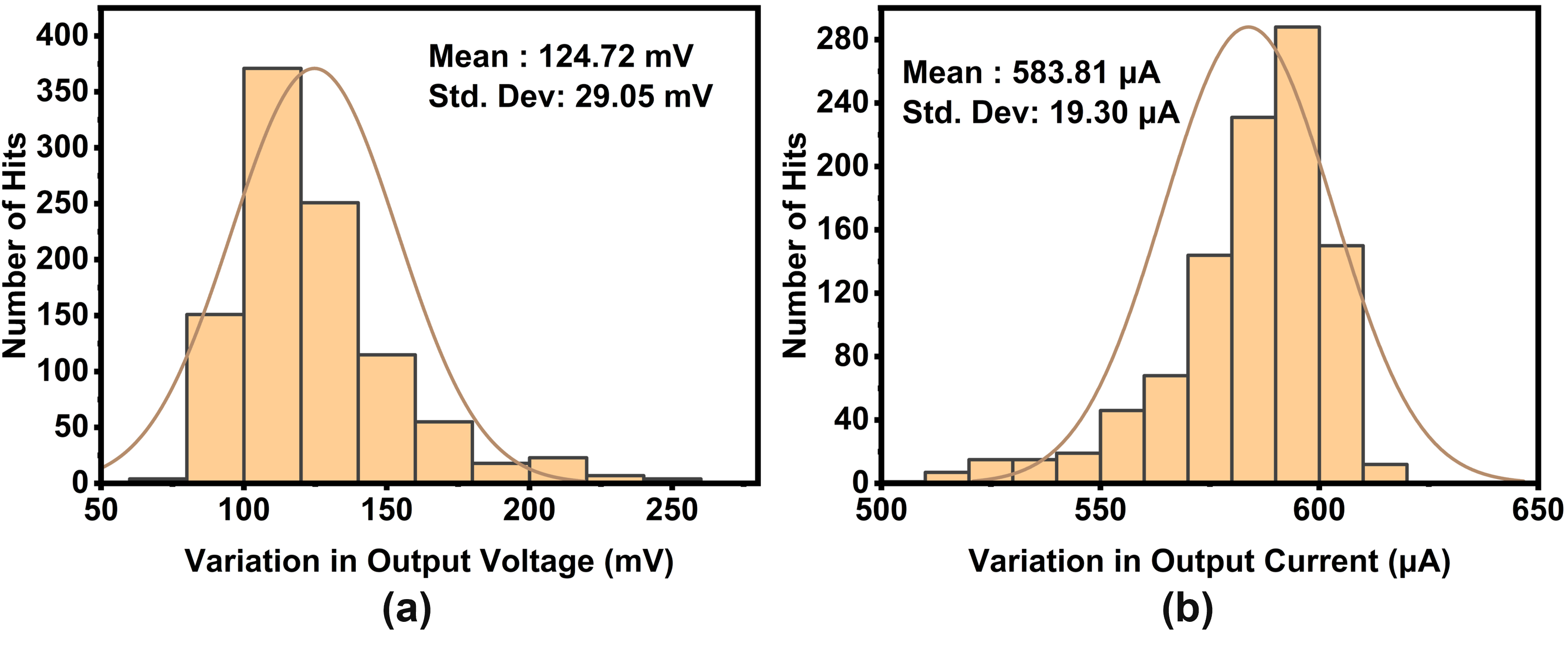}
\vspace{-0.3 in}
\caption{Variation of change in output: (a) Voltage; (b) Current for 128-row activation with weight of 15 during PIM mode.}
\vspace{-0.1 in}
\label{SM}
\end{figure}

The analog output of the proposed $128\times512$ array was digitized using a 6-bit SAR ADC operating at $50~\text{MHz}$. Fig.~\ref{adc_linearity}(a) shows the weight-to-digital transfer characteristics for both calibrated and uncalibrated cases. In the uncalibrated mode, the ADC output spans only codes 7–48 across the 6-bit output range, thereby utilizing less than 70\% of the available dynamic range. This compression not only reduces the effective resolution but also introduces a systematic offset that distorts the weight-to-code mapping. With the optimization of the CDAC reference voltages, the full 6-bit code space (0–63) is exercised across the 4-bit weight range. As shown in Fig.~\ref{adc_linearity}(b) after post-processing, the calibrated response eliminates the offset present in the uncalibrated case and restores a near-ideal linear mapping. Each weight increment corresponds to an average step of $\sim 4$ ADC codes, thereby preserving monotonicity and maximizing quantization efficiency. Fig.~\ref{SM} shows the variation in output voltage and current when 128 rows are activated simultaneously for all 1's in Monte Carlo simulations.  These results highlight that reference tuning and calibration are essential to fully exploit the ADC’s dynamic range when interfaced with the analog outputs of the proposed 6T-2R array.

\begin{figure}[!t]
\centering
\includegraphics[width=1\linewidth]{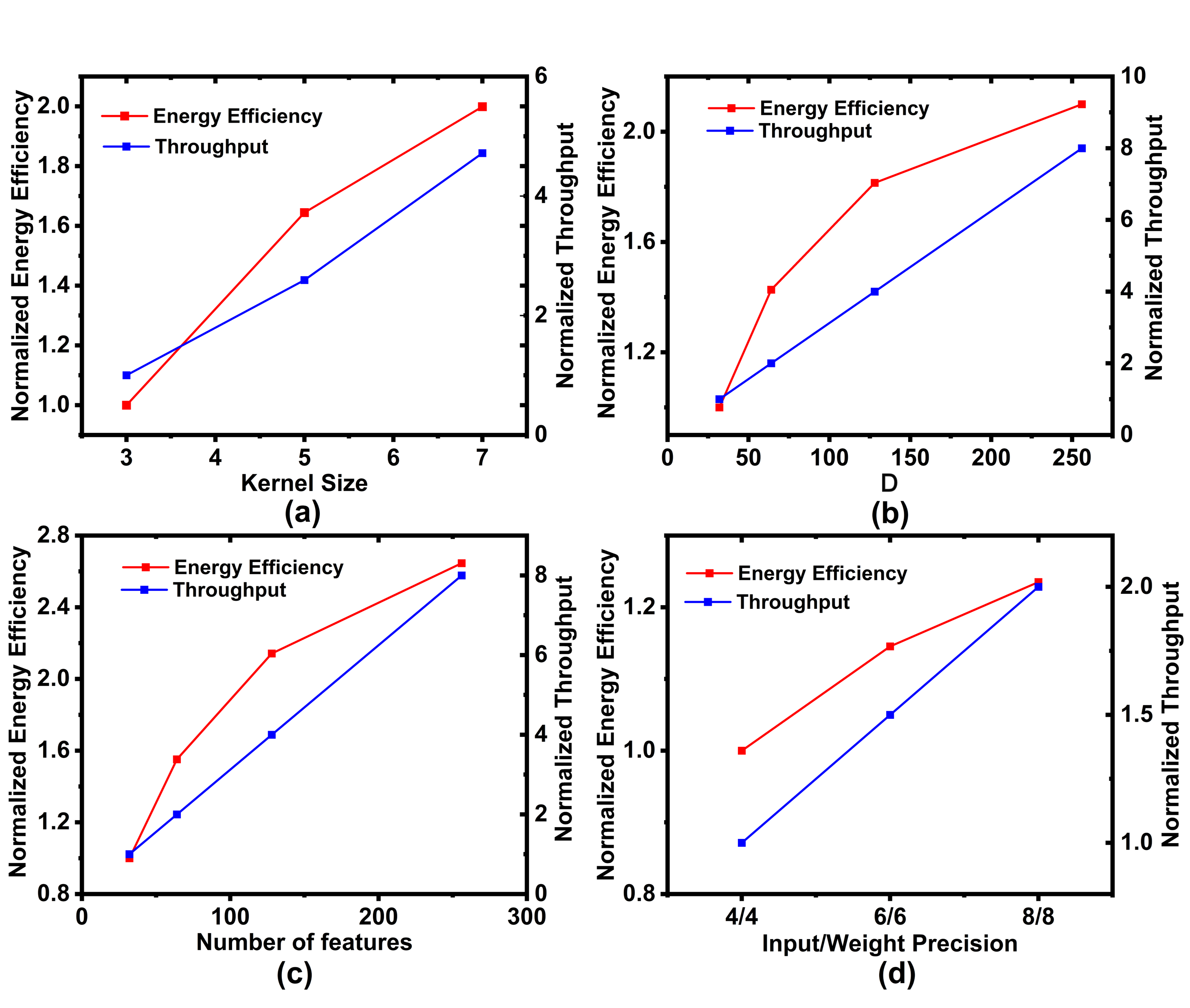}
\vspace{-0.3 in}
\caption{Multi sub-array level evaluation of the proposed 6T-2R PIM: normalized throughput and energy efficiency for 1-bit precision (a) kernel size, (b) parameter D, (c) number of features, and (d) input/weight precision.}
\vspace{-0.1 in}
\label{kernel_system}
\end{figure}

\vspace{-0.1 in}
\subsection{Throughput and Energy Efficiency}
The proposed 6T-2R subarray consists of $128 \times 512$ bit-cells and follows the input-activation and weight mapping strategy described in the previous section, with both input activation and weight precision set to 4 bits. The system latency is dominated by the SAR ADC, which operates at 50~MHz and incurs a delay of 160~ns. Due to the bit-serial computation approach, the latency for performing PIM using $R_{LEFT}$ is 640~ns, and the same applies for $R_{RIGHT}$. After completing two PIM computations, the normalized throughput achieved is 0.4~TOPS at 0.8~V for maximum array utilization. The overall system energy is primarily dominated by the 6T-2R array, which accounts for approximately 60\% of the total energy consumption, followed by the ADC and the WCC block which operates at 1 V. The normalized energy efficiency of the 6T-2R array is calculated to be 452.34~TOPS/W. In terms of area, the total macro size is approximately 0.1~mm$^2$, with the ADC occupying nearly 70\% of the area. The resulting normalized compute density is estimated to be 4.37~TOPS/mm$^2$.

Multiple sub-array level performance of the proposed 6T-2R PIM was evaluated across various neural network parameters, as shown in Fig.~\ref{kernel_system}. Fig.~\ref{kernel_system} (a) illustrates that increasing kernel size improves both throughput and energy efficiency, achieving nearly $2\times$ higher energy efficiency and $\sim1.8\times$ higher throughput at a $7\times7$ kernel compared to the baseline $3\times3$ configuration. This trend results from enhanced parallelism within the array, allowing more MAC operations per access. A similar scaling effect is observed in Fig.~\ref{kernel_system} (b), where increasing parameter depth (D) significantly boosts performance. As the parameter D grows from 32 to 256 layers, throughput improves nearly $8\times$, while energy efficiency more than doubles, demonstrating that the proposed architecture sustains efficiency across wide range of network parameters due to high utilization of the compute elements. Fig.~\ref{kernel_system} (c) shows the effect of increasing the number of features (N). With higher feature dimensions, throughput increases almost linearly due to greater parallel accumulation, while energy efficiency reaches up to $2.7\times$ improvement, confirming that the design effectively exploits large feature maps. Finally, Fig.~\ref{kernel_system} (d) highlights the role of input/weight precision. Increasing precision from 4/4 to 8/8 bits enhances both throughput and energy efficiency. Overall, these results confirm that the proposed 6T-2R NV-SRAM PIM scales favorably with kernel size, depth, input/weight precision and feature dimension. This makes the design highly efficient and adaptable for edge AI workloads where approximate computing is acceptable.  

%\vspace{-0.1in}

\begin{table*}[ht]
\centering
\caption{Comparison Between the Proposed PIM and Previous PIM}
\vspace{-0.1in}
\begin{tabular}{|c|c|c|c|c|c|c|c|}
\hline
\textbf{Parameter} & \textbf{TCASII'24\cite{TCAS2}} & \textbf{ISSCC'23\cite{ISSCC23}} & \textbf{ISSCC'22\cite{isscc22}} & \textbf{TCASI'23\cite{chen2023samba}} & \textbf{TCASI'23\cite{song202328}} & \textbf{JSSCC'24\cite{nverma}} & \textbf{This Work} \\
\hline
\makecell{\textbf{Technology}} & 180nm CMOS & 28nm FDSOI & 22nm FDSOI & 65nm CMOS & 28nm FDSOI & 22nm FDSOI & 22nm FDSOI \\
\hline
%\makecell{\textbf{Supply}\\\textbf{Voltage}} & 0.7-0.9 V & 0.82 V & 0.5-0.8 V & 1.2 V & 0.9 V & 0.8 V & -- \\
%\hline
\makecell{\textbf{Array Size}} & 8Kb & 16Kb & 256Kb & 101Kb & 16Kb & 256Kb & 64Kb \\
\hline
\makecell{\textbf{PIM Domain}} & Time & Charge & Current & Charge & Charge & Current & Current \\
\hline
\makecell{\textbf{Memory Type}} &\makecell{6T SRAM \\ + 9T } & \makecell{10T1C \\ SRAM} & \makecell{1T1R \\ RRAM} &  \makecell{10T1C \\ SRAM} & \makecell{6T \\ SRAM} & \makecell{1T1R \\ MRAM} & \makecell{6T-2R \\ SRAM \\ +RRAM} \\
\hline
\makecell{\textbf{Cache Data} \\ \textbf{Retention}} & No & No & No & No & No & No & Yes \\
\hline
%\makecell{\textbf{NVSRAM}} & No & No & No & No & No & No & Yes \\
%\hline
\makecell{\textbf{Dataset}} & CIFAR-10 & NA & CIFAR-10 & CIFAR-10 & CIFAR-10 & CIFAR-10 & CIFAR-10 \\
\hline
\makecell{\textbf{Accuracy}} & 86.1\% & NA & 91.74\% & 88.6\% & 85.07\% & 90.25\% & 91.76\% \\
\hline
\makecell{\textbf{Input/Weight} \\ \textbf{Precision}} & 8/8 & 8/8 & 8/8 & 8/8 & 4/4 & 4/4 & 4/4 \\
\hline
\makecell{\textbf{Output Precision}} & 14-16${}^{\text{c}}$ & 8 & 19 & 8 & 4 & 6 & 6\\
\hline
\makecell{\textbf{Throughput}\\(GOPS)} & 0.07 & 7.65 & 142.2 & 12.8 & 12.8 & 54.3 & 25.6 \\
\hline
\makecell{\textbf{Energy Efficiency}\\(TOPS/W)} & 0.291 & 16.02 & 0.96 & 10.3 & 16.1 & 5.26 & 28.97 \\
\hline
\makecell{\textbf{Normalized} \\ \textbf{Throughput}\\(TOPS)${}^{\text{a}}$} & 0.2${}^{\text{b}}$ & 0.49 & 5.1 & 3.28 & 0.2 & 0.87 & 0.4 \\
\hline
\makecell{\textbf{Normalized} \\ \textbf{Energy Efficiency}\\(TOPS/W)${}^{\text{a}}$} & 768.7${}^{\text{b}}$ & 1025.2 & 61.8 & 659.2 & 257.6 & 84.2 & 452.34 \\
\hline
\makecell{\textbf{Normalized}\\ \textbf{Compute Density}\\(TOPS/mm${}^{\text{2}}$)${}^{\text{a}}$} & 0.9${}^{\text{b}}$ & 1.19 & 7.9 & 1.52 & 3.59 & 10.9 & 4.37 \\
\hline
\end{tabular}
\vspace{0.5em}

\raggedright
\small{
a. Normalized GOPS or TOPS/W or TOPS/mm${}^{\text{2}}$ = (GOPS or TOPS/W or TOPS/mm${}^{\text{2}}$) × input precision × weight precision, normalized to 1 bit. b. Normalized to 1 bit at 28 nm technology as reported in the paper. c. Utilizes a time to digital converter (TDC).
}

\vspace{-0.1in}
\label{tab:pim_comparison}
\end{table*}

\begin{table}[!t]
\centering
\caption{Comparison of CIFAR-10 accuracy for ResNet-18 under ADC output transfer curves.}
\label{tab:adc_results}
\begin{tabular}{|c|c|}
\hline
\textbf{Configuration} & \textbf{Accuracy (\%)} \\ \hline
Baseline (no ADC nonlinearity or noise) & 92.88 \\ \hline
%ADC nonlinearity + noise (no fine-tuning) & 9.9 \\ \hline
ADC nonlinearity only (fine-tuned) & 92.00 \\ \hline
ADC nonlinearity + noise (fine-tuned) & 91.76 \\ \hline
\end{tabular}
\label{tab:adc_results}
\end{table}

\subsection{Accuracy Analysis}

The influence of non-linear ADC output quantization on CIFAR-10 classification was analyzed using a ResNet-18 architecture. The ADC output employed a 6-bit signed output range, where the original 32-bit floating-point activations were first mapped into this range to emulate quantization. The non-linear behavior of the 6T-2R array was characterized by a curve-fitted polynomial derived from both simulation and SPICE measurements, modeling the transfer characteristics during forward propagation. After the non-linear transformation, the values were inversely mapped back to their original dynamic range.

To emulate hardware-induced variations, we injected Gaussian noise with variable standard deviations estimated from Monte Carlo simulations into the ADC output. This perturbation was applied after the inverse mapping, with its amplitude scaled to the dynamic range of the activation outputs, ensuring that the simulated quantization faithfully captured target hardware conditions. The experimental outcomes are summarized in Table~\ref{tab:adc_results}. The baseline ResNet-18 model achieved an accuracy of 92.88\%. After fine-tuning for 200 epochs with an SGD optimizer (learning rate 0.001, cosine annealing schedule), enabling non-linear ADC quantization without noise reduced accuracy to 92.00\%, and the inclusion of scaled Gaussian noise caused a further drop to 91.76\%. In contrast, without fine-tuning, the model achieved only ${\sim}$77\% accuracy on CIFAR-10, underscoring the necessity of task-aware adaptation when deploying under realistic ADC non-idealities.

\subsection{Comparison with previous works}
%\vspace{-0.1 in}
Table~\ref{tab:pim_comparison} provides a detailed comparison of the proposed 6T-2R NV-SRAM based PIM architecture with state-of-the-art PIM designs. Prior works have primarily relied on time/charge-domain SRAM \cite{TCAS2}, \cite{chen2023samba}, \cite{song202328} or 1T1R RRAM/MRAM arrays~\cite{isscc22,nverma} to demonstrate in-memory computing. In contrast, our work integrates RRAM devices within a conventional SRAM framework, thereby preserving the native bit-cell area and compatibility with standard memory arrays. This makes the proposed bitcell a drop-in replacement for existing SRAM-based cache.  
In terms of throughput and energy efficiency, the proposed design achieves $0.4$~TOPS and $452.34$~TOPS/W when adjusted to 1-bit precision, which is comparable to prior implementations of PIM based on CMOS and ReRAM. The normalization to 1-bit precision follows standard convention in CIM/PIM benchmarking, ensuring fair comparison across different input/output precisions.  It should be noted that the latency bottleneck in our proposed scheme originates primarily from the 6-bit SAR ADC, which operates at $50~\text{MHz}$. While sufficient for functional demonstration, higher-speed ADCs or parallel conversion can significantly improve latency, which would directly increase the achievable throughput (TOPS). Moreover, the ADC also contributes a substantial portion of the reported area. Future solutions, such as 3D integration or ADC sharing across multiple columns or cache banks, can greatly reduce this overhead. Since the proposed 6T-2R bit-cell occupies the same area as a conventional SRAM cell, the overall bit-cell area density remains unaffected, and improvements in ADC design will directly translate into both higher compute density (TOPS/mm$^2$) and higher throughput efficiency (TOPS/W).  
Taken together, these results highlight that while earlier works have shown CIM functionality using charge-domain SRAM or RRAM/MRAM crossbars, our approach uniquely demonstrates SRAM-compatible PIM with embedded RRAM, combining the benefits of area neutrality, non-volatility, and competitive throughput/energy efficiency. This makes the proposed design well-suited as a scalable drop-in replacement for conventional SRAM macros in future cache and memory hierarchies.

\section{Conclusion}
This paper presented an NVM-in-Cache architecture based on a 6T-2R hybrid bit-cell that integrates resistive RAM (RRAM) devices within a conventional SRAM framework. The proposed design doubles the overall memory capacity and enables NVM in-memory computation without altering the native cell footprint, making it a drop-in replacement for standard SRAM cache. By retaining cache data during computation, the architecture eliminates costly cache flush and reload operations while simultaneously enhancing storage density. Circuit-level simulations in 22~nm FDSOI confirm robust resistive switching, with fast 4~ns programming and a high HRS/LRS ratio that ensures reliable binary storage. Furthermore, integration with a calibrated 6-bit SAR ADC achieves full code utilization, maintaining monotonicity and efficient digital mapping of analog weights. System-level evaluations show a throughput of $0.4$~TOPS and an energy efficiency of $452.34$~TOPS/W, demonstrating performance. For 128 row-parallel operations, the CIFAR-10 classification achieves an accuracy of 91.27\%. While the ADC currently dominates both latency and area, future improvements such as 3D stacking and ADC sharing across columns can further boost throughput and compute density. This work does not preclude implementing purely SRAM-based PIM schemes (without NVM); rather, it can be used in conjunction with such prior approaches, effectively doubling the memory capacity and expanding the opportunities to realize IMC within conventional 6T cells. Overall, the proposed 6T-2R NV-SRAM provides a scalable and energy-efficient solution for next-generation cache hierarchies and AI accelerators.

\section*{Acknowledgment }
The research was funded in part by National Science Foundation (NSF) through award CCF2319617.

%\vspace{-0.1 in}
\bibliographystyle{IEEEtran}
\bibliography{ref}

\end{document}